\documentclass[twocolumn,aps]{revtex4}
\usepackage{amsfonts}
\usepackage{amsmath}
\usepackage{amssymb}
\usepackage{graphicx}
\usepackage{color}
\usepackage{ulem}
\usepackage{tabularx}
\usepackage[colorlinks, urlcolor=cyan, citecolor=blue, linkcolor=magenta]{hyperref}

\begin{document}

\title{Mode softening in time-crystalline transitions of open quantum systems%
}
\author{Xiaotian Nie}
\author{Wei Zheng}
\email{zw8796@ustc.edu.cn}
\affiliation{Hefei National Laboratory for Physical Sciences at the Microscale and
Department of Modern Physics, University of Science and Technology of China,
Hefei 230026, China}
\affiliation{CAS Center for Excellence in Quantum Information and Quantum Physics,
University of Science and Technology of China, Hefei 230026, China}
\affiliation{Hefei National Laboratory, University of Science and Technology of China,
Hefei 230088, China}
\date{\today }

\begin{abstract}
In this work, we generalize the concept of roton softening mechanism of spatial
crystalline transition to time crystals in open quantum systems. We study a
dissipative Dicke model as a prototypical example, which exhibits both
continuous time crystal and discrete time crystal phases. We found that on
approaching the time crystalline transition, the response function diverges
at a finite frequency, which determines the period of the upcoming time
crystal. This divergence can be understood as softening of the relaxation
rate of the corresponding collective excitation, which can be clearly seen
by the poles of the response function on the complex plane. Using this mode
softening analysis, we predict a time quasi-crystal phase in our model, in
which the self-organized period and the driving period are incommensurate.
\end{abstract}

\maketitle

\section{Introduction}
\begin{figure}[t]
\includegraphics[width=0.45\textwidth]{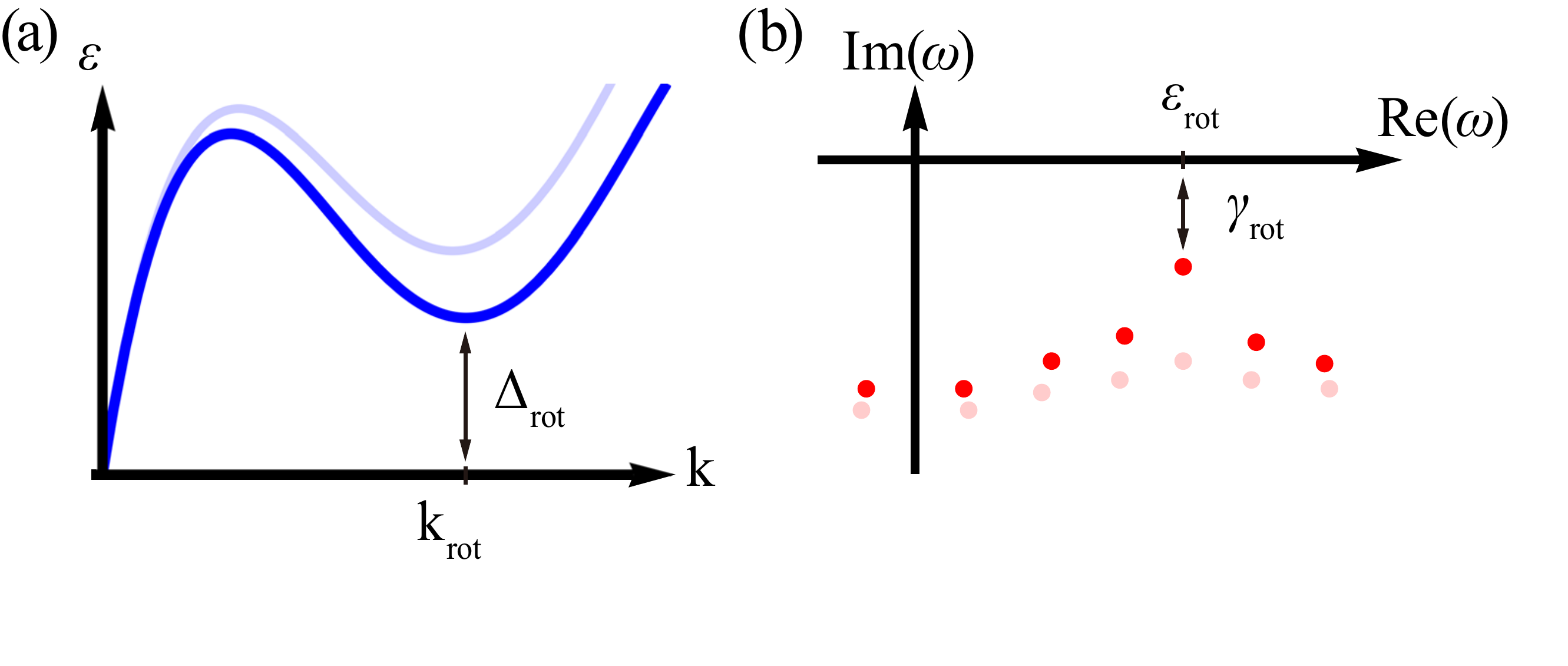}
\caption{Soft modes before the transition to (a) spatial crystals and (b)
time crystals. (a) The excitation spectrum with a roton structure in a quantum
liquid near the transition to a spatial crystal. (b) The poles of the
response function in an open quantum system on the complex plane near the
transition to a time crystal. }
\label{roton}
\end{figure}

Order-to-disorder phase transitions are usually associated with mode
softening. For example, near the transition to a crystal, which breaks the spatial translation symmetry, the excitation spectrum of a quantum
liquid will exhibit a local minimum at a finite momentum $\mathbf{k}_{%
\mathrm{rot}}$ called roton, see Fig.~\ref{roton}(a). As the roton softens,
i.e. the roton gap vanishes, $\Delta _{\mathrm{rot}}\rightarrow 0$, the
quantum liquid becomes unstable, and tends to form a crystal with a period
given by $2\pi/\left|\mathbf{k}_{\mathrm{rot}}\right|$. As a consequence,
when crossing the crystalline transition, the density response function will
diverge at the roton momentum and zero frequency, $\chi \left( \mathbf{k}=%
\mathbf{k_\mathrm{rot}},\omega =0\right) \to \infty$. Roton structure was
first found in the spectrum of superfluid $^{4}$\textrm{He}~\cite%
{roton.He4@Woods.1961,roton@Griffin.1993}. Recently, roton mode softening has also been predicted and observed
in various artificial quantum systems, such as spin-orbit coupled
Bose-Einstein condensates~\cite%
{roton@ZW.2012,roton@Pitaevskii.2012,roton@Engels.2014,roton@Shuai.2015}, superfluids in
shaking optical lattices~\cite{roton@Chin.2015}, quantum gases with dipole-dipole interactions~\cite%
{roton@Lewenstein.2003,roton@Kurizki.2003,roton@Pohl.2010,roton@Ferlaino.2018}, and ultracold atoms coupled with optical cavity~\cite%
{roton@Esslinger.2012}.

Analogues to common crystals, time crystals, which spontaneously break time
translation symmetry, were first proposed by F. Wilczek in 2012~\cite%
{TC@Wilczek.2012,CL-TC@Wilczek.2012}. When the system Hamiltonian is time-independent, it has continuous time translation symmetry. The continuous time crystal (CTC) spontaneously breaks this continuous time translation symmetry, and exhibits permanent periodic oscillation, which is robust against perturbations and the choice of initial conditions. When the system is periodically driven, it has a discrete time translation symmetry. The discrete time crystal (DTC) spontaneously breaks discrete time translation symmetry, and manifests itself as a subharmonic response, which means the system oscillates with $n$ multiple of the driving period for some integer $n>1$.
Soon CTCs were ruled out by the
no-go theorem in the ground states of closed systems~\cite%
{no-go@Bruno.2013,no-go@Oshikawa.2015}. Later, more efforts are put in two
directions. One is to search for the DTC in
periodically driven closed systems~
\cite{FTC@Sacha.2015,DTC@Nayak.2016,Floquet@sondhi.2016,FTC@Sondhi.2016,FTC@Yao.2017(b),FTC@Nayak.2017(b),Rev-DTC@Zakrzewski.2018,Rev-DTC@Yao.2020,DTC@caizi.2021,DTC@Lesanovsky.2019,DTC@caizi.2022}
\cite{DTC@Yao.2017,DTC@Lukin.2017,DTC@Barrett.2018,FTC@Pal.2018,FTC@Nayak.2021,FTC@Yao.2021,FTC@Mi.2022,FTC@Frey.2022,FCT@Stoof.2018}.
 Another way to avoid the no-go theorem is to consider the time crystalline
order in open quantum systems, where dissipation can drive systems into
stable oscillating states~\cite{
OTC@Piazza.2015,OTC@Chan.2015,OTC1@Jaksch.2019,OTC@Lukin.2019,OTC@Cosme.2019,OTC@sacha.2019,OTC@Tuquero.2022,OTC@Jaksch.2020,ODTC@Pal.2020,ODTC@Moessner.2020,OTC2@Jaksch.2019,BTC@Fazio.2018,
BTC@Lesanovsky.2022,DickeCTC@Keeling.2018,KerrLC@Alaeian.2021,CTC@Hemmerich.2019,DTC.Cavity@Ueda.2018,
TC.Cavity@Hemmerich.2020,CTCSC@Esslinger.2022}
. Recently both the DTC and CTC orders have been observed in dissipative atom-cavity systems~\cite{TC.Cavity@Hemmerich.2021,CTC@Hemmerich.2022,OTC@Esslinger.2019,Blue-Cavity@Esslinger.2019,pumping@Esslinger.2022}%
. Compared to spatial crystalline transitions that were driven by roton
softening, a natural question arises, is there a similar mode softening
mechanism in time crystals?

In this paper, we generalize the concept of roton mode softening mechanism to time
crystalline transitions in open quantum systems. We use a modified
dissipative Dicke model as a prototypical example. This model exhibits a CTC
order when the atom-photon coupling is time-independent; while it can enter
a DTC phase as the atom-photon coupling is driven periodically. Using the
Keldysh formalism, we study Gaussian fluctuations in the normal phase near
the transitions to the CTC and DTC phase. We found that the photon response
function, $\chi_{\mathrm{ph}} \left( \omega \right)$ diverges at a finite
frequency $\omega =\varepsilon _{\mathrm{rot}}$ on approaching the time
crystalline transitions. This divergence is controlled by the softening of
the relaxation rate of a collective excitation $\gamma _{\mathrm{rot}%
}\rightarrow 0$, while keeping the excitation frequency to be finite, $%
\varepsilon _{\mathrm{rot}}>0$, during the transition. This frequency $%
\varepsilon _{\mathrm{rot}}$ plays the role of the roton momentum $\mathbf{k}%
_{\mathrm{rot}}$ in spatial crystals, which determines the corresponding
period of the time crystalline orders; while $\gamma _{\mathrm{rot}}$ plays
the role of roton gap $\Delta _{\mathrm{rot}}$, which vanishes at the
transition and leads the normal phase to be unstable. The comparison of the mode softening mechanisms in
spatial crystals and time crystals is given in Table.\ref{tab1}
and Fig.\ref{roton}. This softening mechanism can be clearly seen by the
poles of the response function on the complex plane, where the poles will
cross the real axis at $\varepsilon _{\mathrm{rot}}$ during the transition,
see Fig.\ref{roton}. Using this "roton softening" analysis, we predict a
time quasi-crystal phase in our model, in which the self-organized period
and the driving period are incommensurate~\cite%
{TQC@Volovik.2018,TQC@Sacha.2019}.

This paper is organized as follows. We first introduced the modified Dicke model in Sec. II. Then, we study the phase diagram in the case of a constant atom-cavity coupling and discuss the mode softening in the continuous time crystal by Gaussian fluctuation and exact diagonalization in Sec. III. In Sec. IV, we consider the atom-cavity coupling to be periodically driven, and demonstrate the mode softening in the discrete time crystal and the time quasi-crystal. We summarize with a discussion in Sec. V.

\newcommand{\tabincell}[2]{\begin{tabular}{@{}#1@{}}#2\end{tabular}}
\begin{table}[t]
\caption{Comparison of the mode softening mechanisms in spatial crystals and
time crystals.}
\label{tab1}%
\begin{ruledtabular}
\begin{tabular}{cccccccc}
               &  Spatial crystals                                                 &  Time crystals \\
\hline
Inverse period &  finite $\mathbf{k}_{\mathrm{rot}}$                     &  finite $\varepsilon _{\mathrm{rot}}$  \\
Mode softening      &  $\Delta _{\mathrm{rot}}\to 0$                &  $\gamma _{\mathrm{rot}}\to 0$\\
\tabincell{c}{Divergence\\ of response}     &  $\chi(\mathbf{k}=\mathbf{k}_{\mathrm{rot}},\omega=0)\to\infty$   &  $\chi(\omega=\varepsilon _{\mathrm{rot}})\to\infty$
\end{tabular}
\end{ruledtabular}
\end{table}

\begin{figure*}[htb]
  \includegraphics[width=1.05\textwidth]{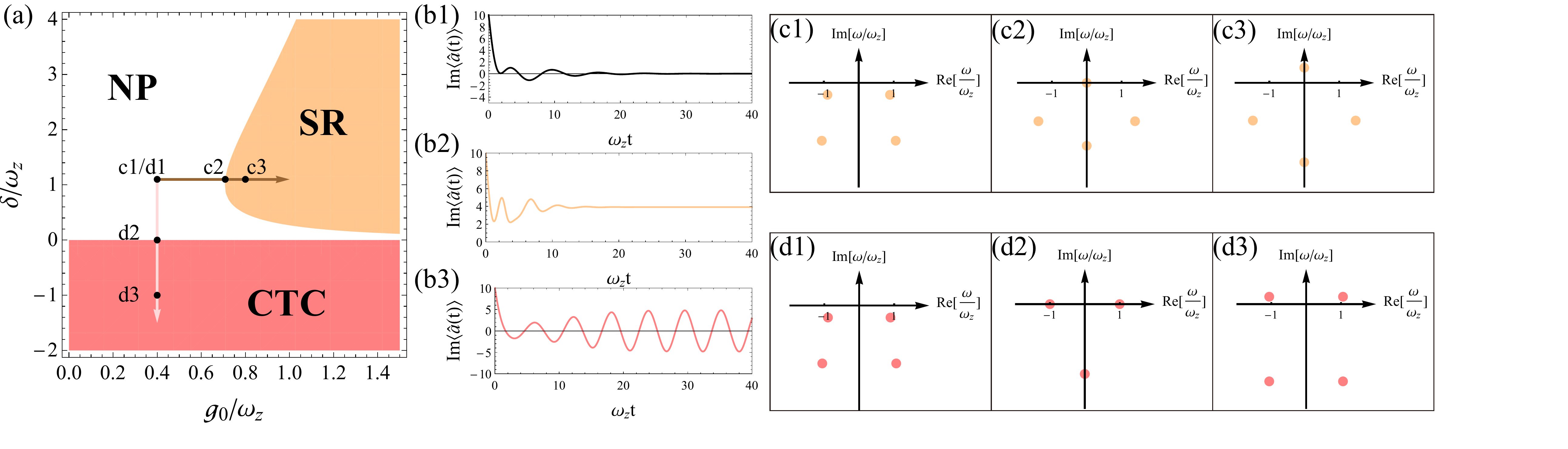}
  \caption{(a) Phase diagram obtained by solving saddle point equations~(\ref{eom1},\ref{eom2}) with parameters $\protect\kappa/\protect\omega_z=1$ and $%
  U/(N\omega_z)=0.01$. (b) Long-time dynamics of three different
  phases. (b1) The normal phase (NP), in which the photon number vanishes
  after a sufficiently long time. (b2) The superradiant phase (SR), in which the system
  reaches a steady state with finite photon occupation. (b3) The continuous
  time crystal phase (CTC). The cavity field oscillates periodically over time and has no steady state. (c)(d) Poles of the response function
  on the complex plane. The corresponding parameters are given by the points
  (c1,c2,c3) and (d1,d2,d3) in the phase diagram Fig.\protect\ref{CTC}(a).}
  \label{CTC}
\end{figure*}

\section{Model}
We consider a modified dissipative Dicke model, which
describes the interaction between $N$ two-level atoms and a single cavity
mode~\cite{Dicke@Keeling.2010,LC@Keeling.2012}. The Hamiltonian takes the
following form ($\hbar =1$),
\begin{eqnarray}
\hat{H} &=&\omega _{0}\hat{a}^{\dagger }\hat{a}+\sum_{i=1}^{N}\frac{\omega _{z}}{2}%
\hat{\sigma}_{i}^{z}  \notag \\
&&+\sum_{i=1}^{N}\frac{g(t)}{\sqrt{N}}\left( \hat{a}+\hat{a}^{\dagger
}\right) \hat{\sigma}_{i}^{x}+\sum_{i=1}^{N}\frac{U}{2N}\hat{a}^{\dagger }%
\hat{a}\hat{\sigma}_{i}^{z},  \label{Ham}
\end{eqnarray}%
where $\hat{a}$, $\hat{a}^{\dagger }$ are the annihilation and creation
operators of cavity photons, and $\hat{\sigma}_{i}^{\alpha }$ with $\alpha
=x,y,z$ are Pauli matrices describing two-level atoms. The cavity frequency
is $\omega _{0}$, level splitting of the atom is $\omega _{z}$, $g(t)$
is the atom-photon coupling, and $N$ is the total atom number. The interaction $U$ can be regarded as a Stark
shift of atomic levels in the cavity field. In the work, we only consider
the situation of $U>0$. Besides the coherent process governed by the
Hamiltonian (\ref{Ham}), leaking of cavity photons leads to dissipative
dynamics.
Thus the system can be described by a Lindblad equation $\partial _{t}\hat{%
\rho}=-i\left[ H,\hat{\rho}\right] +\kappa \left( 2\hat{a}\hat{\rho}\hat{a}%
^{\dagger }-\left\{ \hat{\rho},\hat{a}^{\dagger }\hat{a}\right\} \right) $,
where $\kappa $ is the photon loss rate.

We introduce a collective spin of atoms as $\hat{\mathbf{S}}=\frac{1}{2}%
\sum_{i=1}^{N}\hat{\mathbf{\sigma }}_{i}$. the Hamiltonian can be written into
\begin{eqnarray}
\hat{H} &=&\omega _{0}\hat{a}^{\dagger }\hat{a}+\omega _{z}\hat{S}_{z} +\frac{U}{N}\hat{a}^{\dagger}\hat{a}\hat{S}^{z}\notag \\
&&+\frac{2g(t)}{\sqrt{N}}\left( \hat{a}+\hat{a}^{\dagger
}\right) \hat{S}^{x}.  \label{Ham2}
\end{eqnarray}%
A Holstein-Primakoff
transformation is performed, such that the collective spin can be expressed
by bosons, $\hat{S}_{z}=\hat{b}^{\dagger }\hat{b}-N/2$ and $\hat{S}^{+}=\hat{%
S}_{x}+i\hat{S}_{y}=\hat{b}^{\dagger }\sqrt{N-\hat{b}^{\dagger }\hat{b}}.$
Then we expand the Hamiltonian (\ref{Ham}) to the order of $O(1/N)$ as
\begin{eqnarray}
\hat{H} &\approx& \delta \hat{a}^{\dagger }\hat{a%
}+\omega _{z}\hat{b}^{\dagger }\hat{b}+g(t)\left( \hat{a}+\hat{a}^{\dagger
}\right) \left( \hat{b}+\hat{b}^{\dagger }\right)\nonumber\\
&& -\frac{g(t)}{2N}\left(
\hat{a}+\hat{a}^{\dagger }\right) \hat{b}^{\dagger }\left( \hat{b}+\hat{b}%
^{\dagger }\right) \hat{b}+\frac{U}{N}\hat{a}^{\dagger }\hat{a}\hat{b}%
^{\dagger }\hat{b},
\end{eqnarray}
 where $\delta =\omega _{0}-U/2$.

To investigate the non-equilibrium dynamics, we employ the Keldysh path
integral approach of open quantum systems~\cite%
{Dicke@Diehl.2013,QFT-OS@Diehl.2016}.
The Keldysh path integral is equivalent to the Lindblad master equation of the density matrix. As we know the density matrix can be acted on from both sides, thus there are a time-forward($+$) and a time-backward($-$) components of the fields in the Keldysh formalism~\cite{QFT-OS@Diehl.2016}. By doing the path-integral in the basis of coherent states on the two time branches, bosonic operators are replaced by time-dependent complex-valued fields. The Keldysh partition function is given by
\begin{eqnarray}
Z=\int \mathcal{D} \left[a_{+}^{\ast },a_{+},a_{-}^{\ast },a_-,b_{+}^{\ast },b_+,b_{-}^{\ast},b_-\right]e^{iS},
\end{eqnarray}
and the Keldysh action is
\begin{eqnarray}
S&=&\int_t\{a_{+}^{\ast } i\partial_t a_{+} + b_{+}^{\ast } i\partial_t b_{+} -H_+ \nonumber\\
&&-a_{-}^{\ast } i\partial_t a_{-} - b_{-}^{\ast } i\partial_t b_{-} +H_- \nonumber\\
&&-i\kappa\left(2a_+a_-^\ast - a_+^\ast a_+ - a_-^\ast a_-\right)\},
\end{eqnarray}
in which $H_\pm$ is given by
\begin{eqnarray}
H_\pm&=&\delta {a}_{\pm}^{\ast }{a}_{\pm}%
+\omega _{z}{b}_{\pm}^{\ast }{b}_{\pm}+g(t)\left({a}_{\pm}+{a}_{\pm}^{\ast
}\right)\left({b}_{\pm}+{b}_{\pm}^{\ast}\right)\nonumber\\
&&-\frac{g(t)}{2N}\left({a}_{\pm}+{a}_{\pm}^{\ast }\right){b}_{\pm}^{\ast}\left({b}_{\pm}+{b}_{\pm}^{\ast}\right){b}_{\pm}\nonumber\\
&&+\frac{U}{N}{a}_{\pm}^{\ast }{a}_{\pm}{b}_{\pm}^{\ast }{b}_{\pm}.
\end{eqnarray}

Then we apply the Keldysh rotation,  $\psi_{\mathrm{cl}}=\left(\psi_{+}+\psi_{-}\right)/\sqrt{2},\psi_{\mathrm{q}}=(\psi_+-\psi_-)/\sqrt{2}$, where $\psi=a,b$. The index $\mathrm{cl}(\mathrm{q})$ stands for the 'classical' ('quantum') part of fields. These fields are named 'classical' and 'quantum', only because the former can acquire an finite expectation value while the latter cannot. It does not indicate the 'classical' field can not fluctuate.

In this new basis, we write the total action into $S=S_{2}+S_{4}/N$. Here $S_{2}$ is the quadratic action given by
\begin{eqnarray}
S_{2} &=&\int_{t}\left( a_{\mathrm{cl}}^{\ast },a_{\mathrm{q}}^{\ast
}\right)
\begin{pmatrix}
0 & i\partial _{t}-\delta -i\kappa \\
i\partial _{t}-\delta +i\kappa & 2i\kappa%
\end{pmatrix}%
\begin{pmatrix}
a_{\mathrm{cl}} \\
a_{\mathrm{q}}%
\end{pmatrix}\nonumber\\
&&+\int_{t}\left( b_{\mathrm{cl}}^{\ast },b_{\mathrm{q}}^{\ast
}\right)
\begin{pmatrix}
0 & i\partial _{t}-\omega _{z} \\
i\partial _{t}-\omega _{z} & 0%
\end{pmatrix}%
\begin{pmatrix}
b_{\mathrm{cl}} \\
b_{\mathrm{q}}%
\end{pmatrix}\nonumber\\
&&-4g(t)\int_{t}\left[ \mathrm{Re}\left( a_{\mathrm{cl}}\right)
\mathrm{Re}\left( b_{\mathrm{q}}\right) +\mathrm{Re}\left( a_{\mathrm{q}%
}\right) \mathrm{Re}\left( b_{\mathrm{cl}}\right) \right],
\end{eqnarray}
and the quartic term $S_{4}$ is
\begin{eqnarray}
S_{4} &=&\frac{g}{2}\int_t \left[ \mathrm{Re}\left(a_{\mathrm{cl}}+a_{\mathrm{q}}\right)
\mathrm{Re}\left(b_{\mathrm{cl}}+b_{\mathrm{q}}\right)
\left\vert b_{\mathrm{cl}}+b_{\mathrm{q}}\right\vert^{2}\nonumber\right.\\
&&\left. -\mathrm{Re}\left(a_{\mathrm{cl}}-a_{\mathrm{q}}\right)
\mathrm{Re}\left(b_{\mathrm{cl}}-b_{\mathrm{q}}\right)
\left\vert b_{\mathrm{cl}}-b_{\mathrm{q}}\right\vert^{2}\right]\nonumber\\
&&-\frac{U}{4}\int_{t}\left[ \left\vert a_{\mathrm{cl}}+a_{\mathrm{q}%
}\right\vert ^{2}\left\vert b_{\mathrm{cl}}+b_{\mathrm{q}}\right\vert\right.
^{2}\nonumber\\
&& \left.- \left\vert a_{\mathrm{cl}}-a_{\mathrm{q}}\right\vert
^{2}\left\vert b_{\mathrm{cl}}-b_{\mathrm{q}}\right\vert ^{2}\right].
\end{eqnarray}%

Apply the saddle point approximation~\cite{QFT-OS@Diehl.2016},
\begin{eqnarray}
0&=&\left.\frac{\delta S}{\delta a_{\mathrm{q}}^\ast}\right\vert _{a_{\mathrm{q}}=b_{\mathrm{q}}=0}, \\
0&=&\left.\frac{\delta S}{\delta b_{\mathrm{q}}^\ast}\right\vert _{a_{\mathrm{q}}=b_{\mathrm{q}}=0},
\end{eqnarray}

we obtain
\begin{eqnarray}
i\frac{\partial a_{\mathrm{cl}}}{\partial t} &=&\left( \delta -i\kappa +\frac{U\left\vert
b_{\mathrm{cl}}\right\vert ^{2}}{2N}\right) a_{\mathrm{cl}}\nonumber\\
&&\,+2g\left( t\right) \left( 1-\frac{\left\vert b_{\mathrm{cl}}\right\vert
^{2}}{4N}\right) \mathrm{Re}\left( b_{\mathrm{cl}}\right) ,  \label{eom1} \\
i\frac{\partial b_{\mathrm{cl}}}{\partial t} &=&\left( \omega _{z}+\frac{U\left\vert a_{%
\mathrm{cl}}\right\vert ^{2}}{2N}\right) b_{\mathrm{cl}}\nonumber\\
&&\,+2g\left( t\right) \left( 1-\frac{b_{\mathrm{cl}}\left( b_{\mathrm{cl}}+2b_{\mathrm{cl}}^{\ast }\right) }{4N}\right) \mathrm{Re}\left( a_{%
\mathrm{cl}}\right) .  \label{eom2}
\end{eqnarray}%
Setting $a_{\mathrm{cl}}\left( t\right) =\sqrt{2}\left\langle \hat{a}\left(
t\right) \right\rangle $ and $b_{\mathrm{cl}}\left( t\right) =\sqrt{2}%
\left\langle \hat{b}\left( t\right) \right\rangle $, one could reproduce the
usual mean-field equations of motion, which can describe the colletive
dynamics of this open Dicke model~\cite{Dicke@Keeling.2010,LC@Keeling.2012}.

\begin{figure}[thb]
\includegraphics[width=0.48\textwidth]{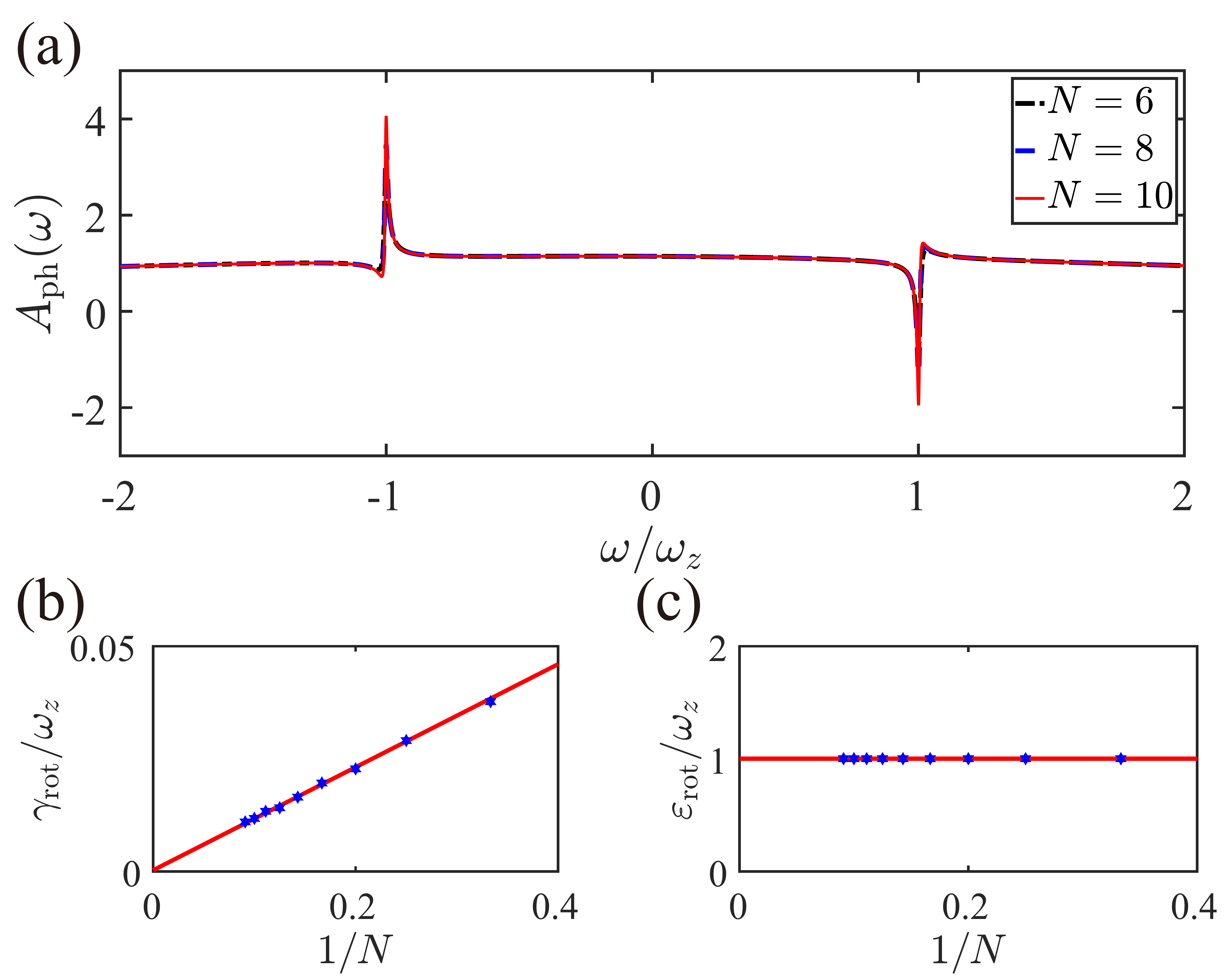}
\caption{(a) Photon spectral functions at $\delta=0$ for finite $N$, in
which $\protect\kappa/\protect\omega_z =3,g_0/\protect\omega_z=0.3,$ and $U/%
\protect\omega_z=0.1$. (b)(c) Finite-size scaling of the damping rate $\protect\gamma_{\mathrm{%
rot}}$ and frequency $\protect\varepsilon_{\mathrm{rot}}$ respectively.
The damping rate is obtained by the full width at half maximum of the peaks, and
the frequency is obtained by the
position of the peaks. The blue dots are the numerical data, and the solid red line is a linear fitting.}
\label{FSS}
\end{figure}

\section{Mode softening in the continuous time crystal}

We first investigate the long-time dynamics for a time-independent atom-cavity coupling
strength $g(t)=g_{0}$. In this situation, the Lindblad master equation has a continuous time translation symmetry. We obtain the phase diagram by solving the saddle
point equations (\ref{eom1},\ref{eom2}), see Fig.\ref{CTC}(a). There are three different
non-equilibrium phases. When $\delta >0$, there is a well-known superradiant
transition at $g_{\ast }=\sqrt{\frac{\delta ^{2}+\kappa ^{2}}{4\delta }%
\omega _{z}}$. If $g_{0}<g_{\ast }$, the system will reach a steady normal
phase (NP) after a sufficiently long time with an empty cavity, $\left\langle \hat{a}\left( t\right)
\right\rangle =0$ (Fig.\ref{CTC}(b1)). As $g_{0}$ exceeds $g_{*}$,
the system will enter a steady superradiant phase (SR). In this phase the cavity
photons condense so that $\left\langle \hat{a}\left( t\right) \right\rangle
\neq 0$, see Fig.\ref{CTC}(b2). In the region of $\delta <0$, no steady
states are found. Instead, the system will develop a CTC order. Both the cavity $%
\left\langle \hat{a}\left( t\right) \right\rangle $ and atoms $\left\langle \hat{b}%
\left( t\right) \right\rangle $ show permanent periodical oscillations in
this phase (Fig.\ref{CTC}(b3)). These oscillations are robust against external perturbations~\cite%
{LC@Keeling.2012}(and see Appendix A). It breaks the continuous time translation symmetry of the
Lindblad equation. In the limit of $\left\vert \delta \right\vert /U\ll 1$, the oscillations are approximately harmonic,
\begin{eqnarray}
\left\langle \hat{a}\left( t\right) \right\rangle &\approx &\frac{2ig_{0}}{%
\sqrt{\omega _{z}^{2}+\kappa ^{2}}}\sqrt{\frac{N\left\vert \delta
\right\vert }{U}}\cos \left( \varepsilon t\right) ,  \label{saddle.point1} \\
\left\langle \hat{b}\left( t\right) \right\rangle &\approx &-\sqrt{\frac{%
N\left\vert \delta \right\vert }{U}}e^{-i\left( \varepsilon t+\phi \right) },
\label{saddle.point2}
\end{eqnarray}%
where $\varepsilon = \omega _{z}+2\left\vert \delta \right\vert
g_{0}^{2}/\left( \omega _{z}^{2}+\kappa ^{2}\right) $, and $\cos \phi
=\kappa /\sqrt{\varepsilon ^{2}+\kappa ^{2}}$.

To investigate mode softening during phase
transitions, we consider quantum fluctuations beyond saddle point
approximation. First, we keep the action to quadratic terms above the
saddle point of the normal phase~\cite{Dicke@Diehl.2013}. It gives $S\approx S_{2}$. Note that in
the thermodynamic limit $N\rightarrow \infty $, $1/N$ terms can be ignored
and this expansion is exact. Here we define a Nambu spinor as $\Psi _{\lambda }(\omega) =\left( a_{\lambda }(\omega)
,a_{\lambda }^{\ast }(-\omega),b_{\lambda }(\omega),b_{\lambda }^{\ast
}(-\omega)\right) ^{\mathrm{T}}$ in frequency domain. $\lambda =\mathrm{cl,q}$, for classical
and quantum components. The quadratic action then can be expressed into a compact form,
\begin{eqnarray}
S_{2} &=&\frac{1}{2}\int_{\omega}\left( \Psi _{\mathrm{cl}}^{\dag }(\omega),\Psi _{%
\mathrm{q}}^{\dag }(\omega)\right)
M_{0}(\omega)
\binom{\Psi _{\mathrm{cl}}(\omega)}{\Psi _{\mathrm{q}}(\omega)},
\end{eqnarray}
where
\begin{eqnarray}
M_{0}(\omega) = \begin{pmatrix}
0 & \left[ G_{0}^{\mathrm{A}}\right] ^{-1} (\omega)\\
\left[ {G_{0}^{\mathrm{R}}}\right] ^{-1} (\omega) & D_{0}^{\mathrm{K}}(\omega)%
\end{pmatrix}%
\end{eqnarray}
and
\begin{eqnarray}
&&\left[G_0^{\mathrm{R}}\right]^{-1}(\omega)=\\
&&\begin{pmatrix}
\omega -\delta +i\kappa & 0 & -g_{0} & -g_{0} \\
0 & -\omega -\delta -i\kappa & -g_{0} & -g_{0} \\
-g_{0} & -g_{0} & \omega -\omega _{z} & 0 \\
-g_{0} & -g_{0} & 0 & -\omega -\omega _{z}%
\end{pmatrix},\nonumber\\
&&\left[ {G_0^{\mathrm{A}}}\right] ^{-1}(\omega)=\left( \left[ {G_0^{\mathrm{R}}}
\right] ^{-1}(\omega)\right) ^{\dag },\\
&&D_0^{\mathrm{K}}(\omega)=2i\kappa \mathrm{diag}\left(1,1,0,0\right).
\end{eqnarray}
The retarded/advanced/Keldysh Green's functions can be obtained by
\begin{eqnarray}
\begin{pmatrix}
G_{0}^{\mathrm{K}}(\omega) & G_{0}^{\mathrm{R}}(\omega)  \\
 G_{0}^{\mathrm{A}}(\omega) & 0%
\end{pmatrix}= M_{0}(\omega)^{-1}.
\end{eqnarray}
One seeks the poles of the retarded Green's function by solving $\det [{G_{0}^{\mathrm{R}}}(\omega
)]^{-1}=0$. Due to the
particle-hole symmetry of Nambu space $\left( \sigma _{x}\otimes I\right) [{G_0^{\mathrm{R}}}(-\omega ^{\ast })]^{-1} \left( \sigma _{x}\otimes I \right) =\left[ {%
G_0^{\mathrm{R}}}(\omega )\right] ^{-1\ast}$, the poles must come in symmetric
pairs about the imaginary axis, see Fig.\ref{CTC}(c,d). The real parts of the poles are the
energies of collective modes. The imaginary parts represent the relaxation
rates of these modes and must be negative. That is to say, the poles
should be always in the lower half complex plane. When a pole happens to
appear in the upper half complex plane, it implies that an excitation mode
will be exponentially amplified in evolution. That will make the system
unstable, and a corresponding phase transition is about to take place.

We analyze trajectories of the poles of response function on the complex
plane near the phase transitions. We found that the superradiant phase
transition and the time crystalline transition occur in fundamentally
different ways. Near the superradiant transition, as $g_{0}\rightarrow
g_{\ast }$ (see trajectory c1-c2-c3 in Fig.\ref{CTC}(a)), a pair of poles
will first move to the imaginary axis, which indicates the vanishing of
excitation energy. After meeting each other on the imaginary axis, this pair
of poles will split in the imaginary direction. Near the transition, one pole is near the axis origin,
\begin{eqnarray}
\omega \approx 2i\sqrt{\frac{\delta(\kappa^2+\delta^2)}{\omega_z \kappa^2}}(g_0-g_*).
\end{eqnarray}
When $g_{0}=g_{\ast }$, the
upper one of the poles will cross the real axis at $\omega =0$, such that
the imaginary part changes its sign to positive~\cite{Dicke@Diehl.2013}, see
Fig.\ref{CTC}(c). That leads to a transition to a steady superradiant phase. Of
course, after the transition, quantum fluctuations above the saddle point of
the normal phase become unstable, and one should analyze the fluctuations
around a correct saddle point, i.e. the superradiant saddle point.

In the other case, in the vicinity of the time crystalline transition ($%
\delta \rightarrow 0^+$), there are a pair of poles near the real axis, see
Fig.\ref{CTC}(d), $\omega =\pm \varepsilon _{\mathrm{rot}}-i\gamma _{\mathrm{%
rot}}$, where%
\begin{eqnarray}
\varepsilon _{\mathrm{rot}} & \approx &\omega _{z}+\frac{2g_{0}^{2}\left(
\omega _{z}^{2}-\kappa ^{2}\right) }{\left( \omega _{z}^{2}+\kappa
^{2}\right) ^{2}}\delta , \\
\gamma _{\mathrm{rot}} & \approx &\frac{4g_{0}^{2}\omega _{z}\kappa }{\left(
\omega _{z}^{2}+\kappa ^{2}\right) ^{2}}\delta .
\end{eqnarray}
These two poles dominate the photon's response function in the long time limit, $\chi_{\mathrm{ph}}(\omega ) = -i\int \mathrm{d}t \, e^{i\omega t}\theta(t) \left\langle \left[ \hat{a}(t),\hat{a}^\dagger(0)\right] \right\rangle$. It can be calculated by the first diagonal element of the retarded Green's function matrix $G_0^{\mathrm{R}}(\omega)$ as
\begin{eqnarray}
\chi_{\mathrm{ph}}(\omega ) \sim  \frac{2\left( \omega +i\gamma
_{\mathrm{rot}}\right) }{\left( \omega +i\gamma _{\mathrm{rot}}\right)
^{2}-\varepsilon _{\mathrm{rot}}^{2}}.
\end{eqnarray}
Note that as $\delta \rightarrow 0^+$,
these two poles approach the real axis, $\gamma _{\mathrm{rot}%
}\rightarrow 0$, but their real parts remain finite $\varepsilon _{\mathrm{%
rot}}\rightarrow \omega _{z}$. At the transition, the poles cross the real
axis at $\omega =\pm \omega _{z}$ instead of $\omega =0$ in superradiant
transition. Thus the response function will diverge at finite frequencies~%
\cite{Diverge@Schir.2019}, $\left\vert {\chi_\mathrm{{ph}} }(\pm \omega _{z})\right\vert \sim 1/\gamma_{\mathrm{rot}}
\sim 1/\left\vert \delta \right\vert $. As a consequence, a time crystalline phase emerges and the system
will oscillate at $\pm \omega _{z}$, right after the transition.  That is
consistent with the previous saddle point solution (\ref{saddle.point1},\ref{saddle.point2}). We
further calculate the photon correlation function, corresponding to the first diagonal element of the Keldysh Green's function matrix $G_0^{\mathrm{K}}(\omega)$, obtaining
\begin{eqnarray}
 C_{\mathrm{ph}%
}(t) &=& \left\langle \left\{ \hat{a}\left( t\right) ,\hat{a}^{\dagger }\left(
0\right) \right\} \right\rangle \nonumber \\
& \approx & \frac{g_{0}^{2}}{\omega _{z}\delta }%
\cos \left( \varepsilon _{\mathrm{rot}}t\right) e^{-\gamma _{\mathrm{rot}%
}\left\vert t\right\vert }
\end{eqnarray}%
near the transition. Note that
both the photon number $\left\langle \hat{a}^{\dagger }\hat{a}\right\rangle =%
\left[ C_{\mathrm{ph}}(0)-1\right] /2= \frac{g_{0}^{2}-\omega _{z}\delta}{2\omega _{z}\delta}$ and the relaxation time $1/\gamma _{%
\mathrm{rot}}$ diverge with exponent $\nu =1$.

To go beyond the Gaussian fluctuation, we
numerically diagonalize the Lindblad equation for a finite atomic number $N$ at the transition point $
\delta =0$, and make a finite-size scaling analysis. We obtain the photon spectral function, $A_{\mathrm{ph}}(\omega
)=-2\mathrm{Im}\chi _{\mathrm{ph}}(\omega )$. The spectral functions of
different numbers of atoms $N$ are plotted in Fig.\ref{FSS}(a). This spectral
function exhibits two peaks around $\omega =\pm \omega _{z}$. We abstract
the damping rate $\gamma _{\mathrm{rot}}$ and excitation energy $\varepsilon
_{\mathrm{rot}}$ from the width and the position of the peaks respectively.
Then we make a finite-size scaling. The results are plotted in Fig.\ref{FSS}%
(b). Note that for a finite atom number $N$, damping rate $\gamma _{\mathrm{%
rot}}$ is finite at $\delta =0$. As $N$ increase, $\gamma _{\mathrm{rot}}$
will approach zero. At the same time, $\varepsilon _{\mathrm{rot}}$ will
remain $\omega _{z}$ as $N\rightarrow \infty $. That is consistent with our
analysis of Gaussian fluctuation.

\begin{figure*}[thb]
  \includegraphics[width=1\textwidth]{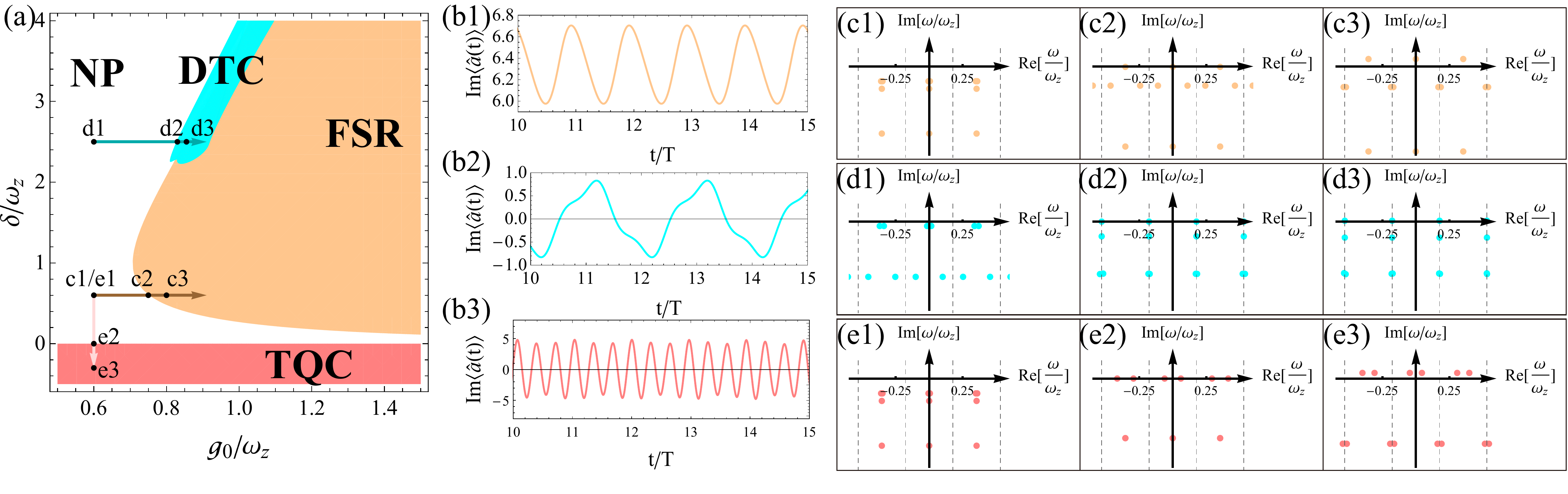}
  \caption{(a) Phase diagram of the periodically driven case, obtained by solving
  saddle point equations~(\protect\ref{eom1},\ref{eom2}) with parameters $\protect\kappa/%
  \protect\omega_z=1$, $g_1/\protect\omega_z=0.05$, $U/(N\omega_z)=0.01$
  and $\Omega/\protect\omega_z=\frac{ \protect\sqrt{2}}{4}\approx0.353$. (b)
  Long-time dynamics of three different phases. (b1) The Floquet superradiant
  phase (FSR), in which the system oscillates with driving period $T$. (b2)
  The discrete time crystal phase (DTC), where the period is doubled, $2T$. (b3) The
  time quasi-crystal phase (TQC). (c)(d)(e) Poles of the response function on
  the complex plane. The dashed lines represent the boundary of
  Floquet-Brillouin zone. The corresponding parameters are given by the points
  (c1,c2,c3), (d1,d2,d3) and (e1,e2,e3), in the phase diagram Fig.\protect\ref{DTC}(a)}
  \label{DTC}
  \end{figure*}

\section{Mode softening in the discrete time crystal and the time quasi-crystal}

In the following, we will demonstrate our mode softening analysis can also be applied to discrete
time crystalline transitions. We modulate the atom-photon coupling strength
periodically $g\left( t\right) =g_{0}+g_{1}\cos \left( \Omega t\right) $. In
this situation, the Lindblad equation is invariant only under a discrete
time translation with period $T=2\pi /\Omega $. By solving the corresponding
saddle point equations (\ref{eom1},\ref{eom2}), we found three different phases at $%
\delta >0$, see Fig.\ref{DTC}(a). When $g_{0}$ is small, the cavity reaches
a steady NP after a sufficiently long time $\left\langle \hat{a}(t)\right\rangle =0$.
When $g_{0}$ is sufficiently large, the system approaches a Floquet
superradiant phase (FSR), in which the cavity field is nonzero and
oscillates with the driving period, see Fig.\ref{DTC}(b1). Between the NP
and the FSR, a DTC phase is found, in which the oscillation period is
doubled (Fig.\ref{DTC}(b2)), thereby breaking the discrete time translation
symmetry.

\begin{widetext}
As before, we consider Gaussian fluctuations in the NP. In contrast to the undriven case, we obtain infinite poles on
the complex plane~\cite{FloquetOQS@Diehl.2019,FloquetOQS@Diehl.2020}, see in Fig.\ref{DTC}(c,d,e). For a periodic potential, different momentum components separated by the reciprocal vector are coupled due to lattice scattering. Quasi-momentum defined in the first Brillouin zone is a good quantum number. Similarly, different frequency components of the same quasi-energy are coupled in our mono-chromatically driven system, the action expressed in the frequency domain is tri-diagonal,
\begin{eqnarray}
   S_2 &=&\frac{1}{2}\int_{-\frac{\Omega}{2}}^{\frac{\Omega}{2}}\mathrm{d}\omega\,
    \begin{pmatrix}
     \cdots  &  \Psi(\omega)^\dagger, &  \Psi(\omega+\Omega)^\dagger & \cdots
    \end{pmatrix}
    \begin{pmatrix}
     \ddots & \ddots &   &   \\
     \ddots & M_{0}(\omega) & M_{1} & \\
      & M_{1} & M_{0}(\omega+\Omega) & \ddots\\
      & & \ddots&\ddots\\
    \end{pmatrix}
    \begin{pmatrix}
     \vdots \\  \Psi(\omega) \\  \Psi(\omega+\Omega) \\ \vdots
    \end{pmatrix}.
\end{eqnarray}
where
\begin{eqnarray}
M_{1} = \begin{pmatrix}
0 & \left[ G_{1}^{\mathrm{A}}\right] ^{-1} \\
\left[ {G_{1}^{\mathrm{R}}}\right] ^{-1}  & 0%
\end{pmatrix},%
\end{eqnarray}
and
\begin{equation}
\left[ {G_1^{\mathrm{R}}}\right] ^{-1}= \left[ {G_1^{\mathrm{A}}}\right] ^{-1} = \frac{1}{2}%
\begin{pmatrix}
0 & 0 & -g_1 & -g_1 \\
0 & 0 & -g_1 & -g_1 \\
-g_1 & -g_1 & 0 & 0 \\
-g_1 & -g_1 & 0 & 0%
\end{pmatrix}.%
\end{equation}%
Such that the dimension of the retarded Green's function $%
G^{\mathrm{R}}\left( \omega \right) $ is infinite rather than $4\times 4$ in
the undriven case
\begin{equation}
\left[ {G^{\mathrm{R}}(\omega )}\right] ^{-1}=
\begin{pmatrix}
     \ddots & \ddots &   &   &   \\
     \ddots & \left[ {G_0^{\mathrm{R}}(\omega-\Omega )}\right] ^{-1} &\left[ {G_1^{\mathrm{R}}}\right] ^{-1}  & & \\
      &  \left[ {G_1^{\mathrm{R}}}\right] ^{-1}  & \left[ {G_0^{\mathrm{R}}(\omega )}\right] ^{-1} &\left[ {G_1^{\mathrm{R}}}\right] ^{-1}  & \\
      & &\left[ {G_1^{\mathrm{R}}}\right] ^{-1}  & \left[ {G_0^{\mathrm{R}}(\omega+\Omega )}\right] ^{-1} &\ddots\\
      & & & \ddots&\ddots\\
    \end{pmatrix},\label{InvFloquetGR}
\end{equation}
\end{widetext}

To calculate the poles of the retarded Green's function numerically, we have to make a dimensional cutoff. In practice, we take 19 Floquet-Brillouin zones into account, i.e. the matrix size is 76$\times$76. Poles manifest perfect periodicity in the five Brillouin Zones nearest to zero and are convergent with growing cutoff size, which means our cutoff is sufficiently large.

The real parts of the poles are equally spaced by $\Omega $. At the
transition to the FSR phase, a chain of poles will cross the real axis at
the center of the Floquet-Brillouin zones, $\omega =n\Omega $, see Fig.\ref{DTC}(c). That means the oscillation period of the upcoming FSR is just the
driving period $T$. When approaching the DTC transition, see Fig.\ref{DTC}%
(c), the chains poles will cross the real axis at the boundary of
Floquet-Brillouin zones, $\omega =\left( n+1/2\right) \Omega $. The photon's
response function will diverge at these frequencies, $\omega =\left(
n+1/2\right) \Omega $. As a consequence of this mode
softening, the oscillation period is doubled after the transition.

When $\delta \rightarrow 0$, we found that the poles will cross the real
axis at $\pm \omega _{z}+n\Omega $. According to our mode softening
analysis, a time crystalline order may emerge at frequency $\pm \omega
_{z}+n\Omega $. If the external driving frequency $\Omega $ and the
intrinsic energy $\omega _{z}$ are incommensurate, the oscillation will
become quasi-periodic. That is to say, a time quasi-crystal(TQC) may emerge in
the $\delta <0$ regime. Thus we numerically solve the saddle point equations
in this regime and plot the long-time evolution in Fig.\ref{DTC}(b3). Note
that the oscillation is quasi-periodic and will not repeat itself in a
finite time. This numerical result is consistent with our mode softening
analysis. The robustness of the DTC and the TQC against perturbations is checked in Appendix A.

\section{Summary and Outlooks}
We generalize the "roton" mode softening
mechanism of spatial crystals to time crystals in open quantum systems. In
time crystalline transition, the softening mechanism is that the damping
rate of a collective mode will vanish, while the energy of this mode remains finite. That indicates the emergence of an undamped mode with non-zero
energy in open systems, which will compete with the existing steady state,
leading to the possible order in the time domain.

In experiments, the Dicke model discussed in
this work can be regarded as a simplified mode of the current
experiments~\cite{CTC@Hemmerich.2022,Blue-Cavity@Esslinger.2019}. The two
internal atomic levels are simulated by the center-of-mass states of atoms, and the coupling $g$ can be tuned via modulating the transversal pumping laser.
The response of atoms can be measured by a Bragg-like probe~\cite%
{roton@Esslinger.2012}, and the correlations of the cavity field can be
obtained by measuring the photons leaking out of the cavity~\cite%
{TC.Cavity@Hemmerich.2021}. We expect that this mode softening in time
crystals can be observed in those experiments.

In this work, our discussion
is limited in open quantum systems. The damping of collective modes is dominated by
external dissipation. However, this mode softening mechanism can be also generalized to closed systems, where the relaxation is induced
by multi-mode couplings.

\textit{Acknowledgment}. We thank Hui Zhai, Zheyu Shi and Junsen Wang for
helpful discussions.

\appendix

\begin{figure}[t]
\includegraphics[width=0.35\textwidth]{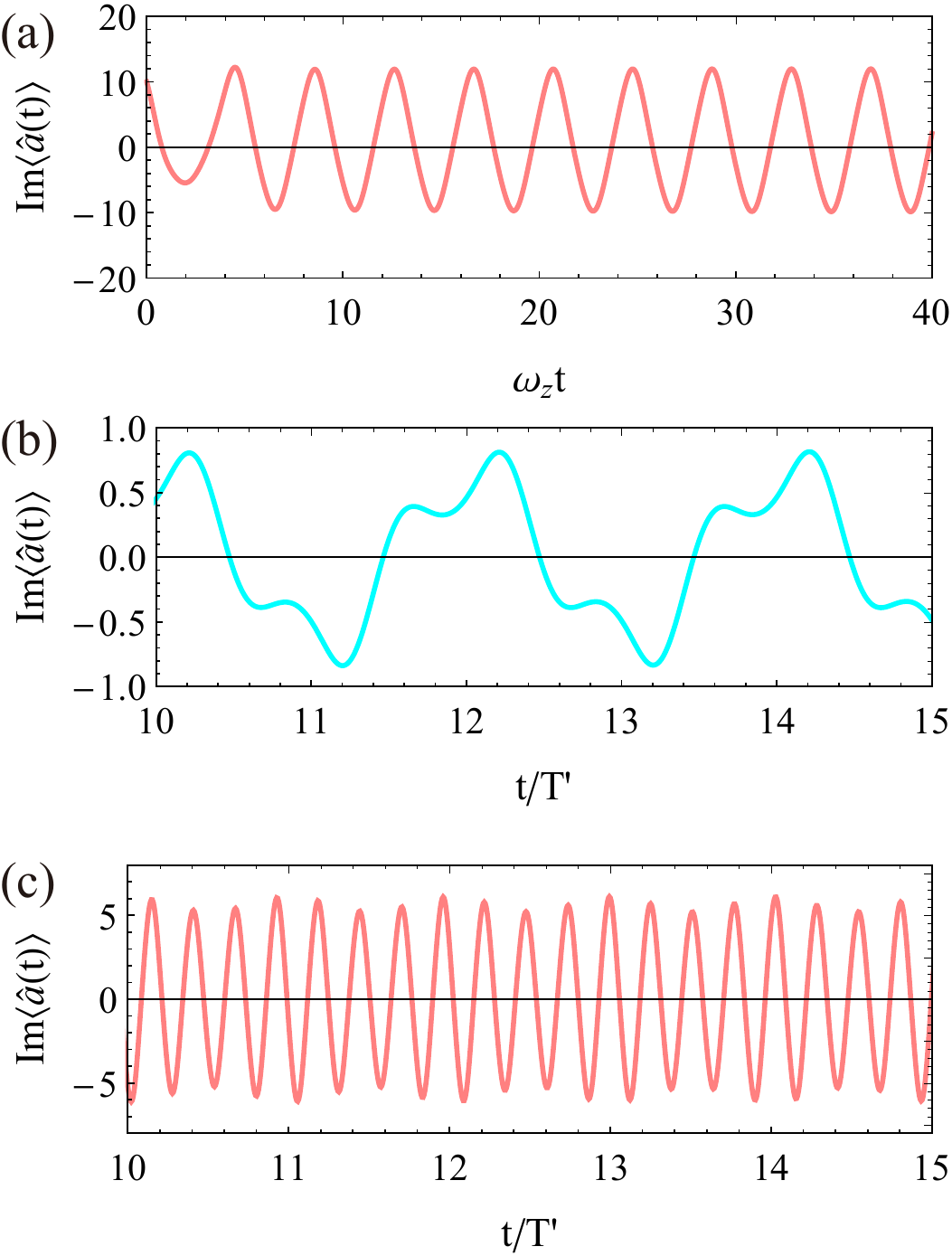}
\caption{Photon dynamics under perturbation in the generalized Dicke model. (a)The CTC phase, (b)The DTC phases and (c) the time quasi-crystal phase. Here $\delta g/g = 0.1$ and the driving frequency is tuned to be $\Omega^\prime/\omega_{z}  = \sqrt{2}/5$.}
\label{robustness}
\end{figure}

\section{Robustness of the Dicke Time Crystals}
Time crystals are robust against perturbations. In this appendix, We will show the time crystal phases, including CTC, DTC and time quasi-crystal, are robust against adding perturbation and changing parameters. We add a perturbation like $\frac{2\delta g}{i\sqrt{N}}(\hat{a}-\hat{a}^\dagger)\hat{S}_y$ into the Hamiltonian~\ref{Ham2}. This perturbation appears as the rotating-wave term and antirotating-wave term of the atom-light coupling are tuned imbalanced. Then we consider the saddle point solutions with $\delta g/g = 0.1$, and investigate the sufficient long time behavior. We found that the CTC phase is robust against the perturbation, see Fig.\ref{robustness}(a).

For the periodically driven case, we remain the unbalanced perturbation $\delta g/g = 0.1$, and tune the driving period to be $\Omega^\prime/\omega_z=\sqrt{2}/5$. In the DTC phase, the system oscillates with a doubled period of driving, $T^\prime=2\pi/\Omega^\prime$, see Fig.\ref{robustness}(b). Besides, the system will again enter the time quasi-crystal phase, when $\delta<0$ (Fig.\ref{robustness}(c)). That indicates the both DTC and time quasi-crystal are robust against such perturbation.

\end{document}